# An Application Package Configuration Approach to Mitigating Android SSL Vulnerabilities

Vasant Tendulkar
Department of Computer Science
North Carolina State University
*tendulkar@ncsu.edu*

William Enck
Department of Computer Science
North Carolina State University
*enck@cs.ncsu.edu*

*Abstract*—Computing platforms such as smartphones frequently access Web content using many separate applications rather than a single Web browser application. These applications often deal with sensitive user information such as financial data or passwords, and use Secure Sockets Layer (SSL) to protect it from unauthorized eavesdropping. However, recent studies have confirmed a wide-spread misconfiguration of SSL verification in applications. This paper considers the difficulty faced by Android application developers when modifying SSL code for using common features like *pinning* or using a self-signed SSL certificate. For example, developing an application that accesses a test Web server with a self-signed certificate requires additional code to remove SSL verification; however, this code is not always removed in production versions of the application. To mitigate vulnerabilities introduced because of the complexity of customizing SSL code in Android applications, we propose that common SSL configuration should be specified in the application's package manifest. We provide two concrete suggestions: 1) **linking the application's debug state to SSL verification, and** 2) **pinning certificates and CAs in the manifest.** We evaluate the appropriateness of these two suggestions on over 13,000 applications from Google's Play Store, of which 3,302 use SSL in non-advertisement code, and find that 1,889 (57.20%) of these SSL applications would benefit.

## I. INTRODUCTION

Smartphones are increasingly integral to our daily lives. A key feature of smartphones is their ability to run third party applications, often downloaded from application markets, such as the Apple App Store and Google Play Store. These applications frequently do not execute in isolation. Instead, they communicate with Web servers to retrieve information and perform computation. If this communication is not secured, network adversaries can eavesdrop on sensitive information such as passwords and financial data, or actively intercept and modify the communication to exploit client software.

The Secure Sockets Layer (SSL) is the *de facto* standard for securing communications over the Internet. SSL and the related Transport Layer Security (TLS) standard have undergone many revisions to ensure a robust security protocol. However, the security of SSL is contingent on correct client-side verification of 1) the server certificate and 2) the server hostname. This logic is controlled by the client application and not the SSL library. For example, the hostname must be extracted from the URL and compared with the common name in the SSL certificate. Failing to verify either the certificate or the hostname enables Man-in-the-Middle (MITM) attacks wherein an active adversary intercepts the SSL connection with the client and relays traffic through a new SSL connection with the server. While all traffic is in fact encrypted, the adversary can easily eavesdrop and modify the contents.

Until relatively recently, the wide-spread end-user software using SSL was mostly limited to Web browsers and Email clients. There were only a handful of such applications, allowing for scrutiny that led to the discovery and removal of verification flaws [23], [24], [25]. The smartphone's "application-centric" computing model creates a new problem. The SSL verification logic has moved from the Web browser application to thousands of applications written by a diverse set of developers. The available scrutiny has remained relatively constant and must be distributed across these thousands of applications. Therefore, many applications with vulnerable SSL verification logic potentially exist. Fahl et al. [16] recently confirmed this expectation for Android applications. They found 41 of 100 manually audited applications using SSL were vulnerable to MITM attacks.

A simple solution to widespread SSL verification flaws is to encapsulate the logic into an easy to use Application Programming Interface (API). Indeed, this is what Android provides. In most cases, the difference between using HTTP and HTTPS to access a Web server is using the `https://` prefix scheme in place of `http://`, and the Android libraries ensure correct verification. The verification flaws observed by Fahl et al. occur when the developer decides to add extra code to define custom SSL socket factories, trust managers, and hostname verifiers that effectively bypass the verification.

In this paper, we propose an application package configuration approach to mitigating Android SSL vulnerabilities. We observe that many of the existing SSL vulnerabilities occur *because the developer needed to add code*. The significant number of developers adding SSL-related code implies that Android's easy to use SSL API interface is insufficient, because it does not enable common features required by developers. By moving SSL configuration to the application's package manifest, we seek to eliminate the need for the developer to add code that will potentially lead to vulnerabilities.

We make the following two suggestions. First, the application's "debug" flag should control SSL verification. If the application is in *debug mode*, then SSL should allow any certificate and hostname. This suggestion accounts for the common case where developers are using test servers that use self-signed certificates. Currently, this scenario requires additional code in development, and the code may not be removed in the production version. Second, the package manifest should allow the application's SSL connections to be "pinned" to a specific set of certificates and certificate authorities (CAs). This suggestion allows developer to $a)$ use production servers with self-signed certificates or certificates not yet trusted by Android, $b)$ limit exposure to vulnerable CAs [8], [12].

We evaluate the usefulness of these two suggestions by surveying a snapshot of 13,000 popular free applications from Google's Play Store and 240 open-source applications from the F-Droid repository [15]. Of the 3,302 applications that use SSL within the application (not counting SSL for advertisements and analytics), we conservatively find that 1,889 applications (57.20%) could have been supported using our package manifest solution.

This paper makes the following contributions:
- We propose that SSL verification logic should be stated in an application's configuration and not in code. We provide two valuable suggestions for doing so.
- We evaluate our two suggestions on a corpus of 3,302 closed-source applications that use SSL within the application. We demonstrate that 57.20% of these applications would benefit from implementing our suggestions.

Finally, we note that moving all potential SSL configuration to the package manifest is not practical or desirable. Providing complete programmability of SSL verification in the package manifest would require a Turing complete sub-language. Rather, our goal is to provide an alternative to adding code that covers a greater majority of use cases. We strongly believe that there will always be edge cases, and removing code-based SSL configuration ability from developers, as suggested by a recent concurrent study by Fahl et al. [17], is overly limiting.

The remainder of this paper proceeds as follows. Section II overviews SSL and its usage in Android. Section III describes our proposed solutions to mitigate SSL vulnerabilities in Android applications. Section IV describes the methodology and the experiments. Section V discusses additional topics. Section VI describes related work. Section VII concludes.

## II. BACKGROUND AND MOTIVATION

### A. Background

**SSL Handshake Protocol:** The client and server establish a secure connection by using the SSL handshake protocol, summarized in Figure 1.

The two steps in the protocol necessary to ensure the authenticity of the connection are verifying the server certificate (chain-of-trust establishment) and verifying the server identity (hostname verification). Since these are implemented in the application layer, developers have the responsibility of implementing them correctly in the application.

**Certificate Verification:** Every SSL client has a list of trusted root CA certificates. Each SSL certificate contains an "issuer" field that has the name of the issuer Certificate Authority (CA).

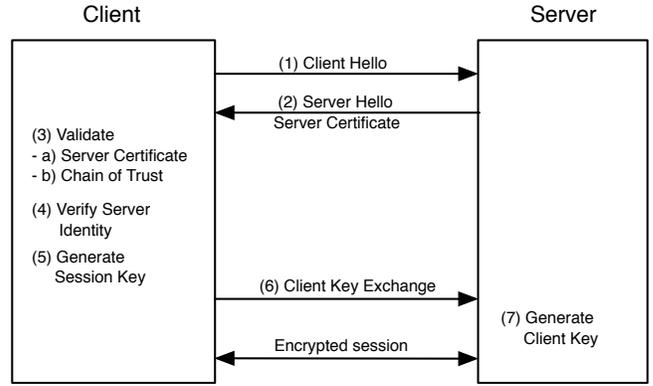

Fig. 1. Overview of the SSL Handshake Protocol

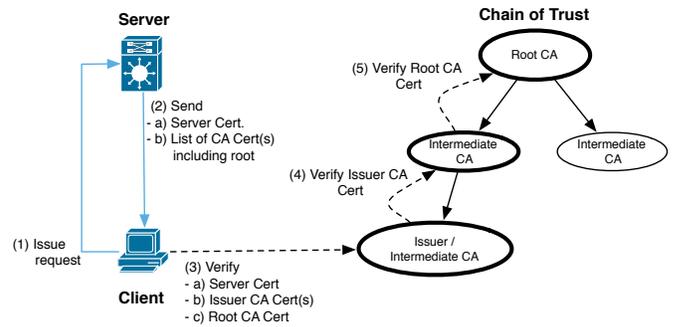

Fig. 2. Establishing the Chain of Trust

In response to the client hello, the server sends its own SSL certificate and the SSL certificate of its issuer CA to the client. If the issuer CA is not a root CA then the server sends a chain of intermediate CAs, all the way to a root CA. The client then attempts to establish a chain-of-trust starting with the server's certificate all the way up the chain to the root CA. The client also verifies that the certificates are not expired and that each CA certificate has the 'CA' bit set in the "Basic Constraints" field. If either the root CA is not trusted by the client or one of the intermediate CA's certificate is either expired or invalid then the client closes the SSL connection.

**Hostname Verification:** Once the chain-of-trust is established, the client then verifies the server's identity (hostname). The client constructs a list of acceptable identifiers based on the server's domain name. RFC 2818 [26] advises to use 'SubjectAltName (SAN)' extension as the main source for server identity. Otherwise, the (most specific) Common Name (CN) field in the 'Subject' field of the certificate must be used. CN use is now deprecated and should be only checked to ensure backward compatibility. After creating a list of server identifiers, as specified in RFC 6125 [27], the client attempts to match the Fully Qualified Domain Name (FQDN) of the server to one of the identifiers.

Usually a single SSL certificate is used per domain. However, if a developer wants to use the same certificate for multiple domains or sub-domains, she would have to use either a SAN enabled certificate for multiple domains or a *wildcard* certificate, both of which are more expensive than regular SSL certificate. SAN certificates can be used on either multiple domains or sub-domains. Alternatively, if the developer wants to use a single certificate on multiple sub-domains of a particular domain then she could use either a wildcard certificate or write custom verification code to accept the same certificate over multiple sub-domains. There are certain vulnerabilities associated with incorrect sub-domain hostname verification while using wildcard certificates. For example, if the developer uses a wildcard certificate for *.foo.com and has custom verification code to accept all hostnames matching *.foo.com without correctly validating the hierarchy, then the attacker's certificate for *.xyz.foo.com could pass as a correct certificate, thus making the application vulnerable to MITM attacks. Hence, irrespective of the certificate used the developer needs to ensure that the implementation for certificate verification code is correct. Matching identifiers containing wildcards is by itself fairly complex. The default hostname verifier used by the Android SDK matches only sub-domains in the same level and not with deeper sub-domains. Developers need to write their own hostname verification code to facilitate hostname verification at multiple sub-domain levels.

After successful completion of the handshake, a key is exchanged between the client and the server, which is used for encrypting the communications between them.

### B. Motivation

There are several cases where developers need to modify the default certificate verification logic to use SSL in their application. For example, if the developer wants to trust only one root CA certificate then she has to modify the default SSL verification logic to pin the root CA certificate to the application i.e. *SSL pinning*. Likewise, if the developer wants to use a self-signed certificate during development or on production servers, then she has to alter the SSL verification logic to consider the self-signed certificate as trusted. We now discuss the ways in which developers can modify default verification logic to use SSL correctly in their application.

The Android SDK provides APIs for developers to create SSL connections, but these built-in APIs only trust the root CA certificates that come bundled with Android. However, the use of the CA model for SSL verification implies that if developers choose to trust the root CA then they also have to trust all the intermediate CAs signed by that root CA. If the developer does not want to trust all the built-in root CA certificates, then she can use SSL pinning to trust only certain root CAs. SSL Pinning is a technique that allows developers to protect the application from MITM attacks resulting from fraudulently issued certificates or compromised CA credentials. To protect the application from compromised CAs, the developer can add custom certificate verification code which discards all the other CA certificates except the pinned ones.

The default SSL verification logic of Android also does not accept self-signed certificates or certificates signed by a CA not currently trusted by Android (untrusted CA). Such certificates can be obtained for free or at minimum cost to the developer. Using self-signed certificates can also protect the application from MITM attacks resulting from compromised CAs. Now the developer has to trust only those certificates that were created and signed by her. Since application updates are easily distributed through app markets, managing self-signed or untrusted CA certificates via pinning is also reasonable. In order to use such certificates, the developer has to modify the default certificate verification logic to trust the self-signed certificate or untrusted CA certificate. This can also be achieved by pinning the aforementioned certificate in the application.

Developers also often use the same certificate or self-signed certificates on multiple development servers to avoid using costly SSL certificates for debugging purposes. However, the production servers use valid SSL certificates. In order to use self-signed certificates, developers can customize the default verification logic to pin them to the application. However, once debugging is complete developers need to remove all traces of the such code to ensure correct verification of SSL certificates in the final production version of the application.

When developers use built-in APIs to create SSL connections server certificate and hostname verification is performed by Android. The connection is closed if either of the verification fails. However, when developers customize SSL connections for any of the above mentioned reasons, they have to modify the default SSL verification logic in the application. Failing to correctly verify either the server certificate or its identity leaves the application vulnerable to MITM attacks. We believe that developers should not be burdened with the responsibility of adding extra code for customizing SSL connections to use a particular kind of certificate or to use popular SSL features. To alleviate this burden from the developers, we provide configuration policy based solutions in Section III and evaluate their effectiveness in Section IV.

### III. SOLUTION

Recent research [16], [20] highlighted some flaws in the use of SSL in mobile and non-browser software. They showed that the incorrect use of SSL in applications and inconsistent implementation of APIs in various libraries leaves applications vulnerable to MITM attacks. Concurrent to our work, Fahl et al. [17] have provided solutions to address these problems. Section VI further differentiates our work from Fahl et al.

To provide effective solutions, we looked at the reasons why developers add custom SSL verification code to applications. In January 2013, we studied programming-related Q&A forums (e.g. StackOverflow.com) for "*using SSL in Android applications*". The following list summarizes the most common causes we found in the replies:

- The developer is using a self-signed certificate.
- The developer is using a certificate signed by a CA not yet trusted by Android.
- The developer does not know how to use SSL and wants to get rid of the `SSLException`.

```
<application
    android:allowBackup="true"
    android:icon="@drawable/ic_launcher"
    android:label="@string/app_name"
    android:theme="@style/AppTheme"
    android:debuggable="true">
```

Fig. 3. Enabling debug flag in Android Manifest

- The developer is using the same certificate on multiple servers.

The solutions provided to most of the questions mainly contained explanations on techniques to accept all certificates, which essentially translates to disabling SSL verification. These answers were commonly called *Allow-All*, *Accept-All* or *Trust-All* solutions (cf. Section IV-C). Very few entries mentioned the negative consequences of writing such code. Disabling SSL verification leaves the application vulnerable to MITM attacks. Studying these entries confirmed our belief that the complexity of customizing SSL verification code to use common features was a substantial factor in driving developers towards using a *Trust-All* or *Allow-All* solution.

Providing new classes with more complex functionality for developers to learn and use is not the appropriate solution. A better solution is to make the process of customizing SSL verification transparent to the developer. We believe that customizing SSL verification should not require adding code to the application, as it can lead to vulnerabilities. We present configuration-based solutions that can be implemented as policies in the application manifest file.

It is possible that there are additional reasons why developers add custom verification code to applications and we might not have been able to cover all possible causes above. There will always be edge cases. However, our goal is to provide an easier alternative to adding code, that covers a majority of the applications, and at the same time giving developers the flexibility to customize SSL connections. We discuss this further in Section V.

### A. Linking application debugging with SSL verification

The manifest file (*AndroidManifest.xml*) in Android applications uses a tag (*android:debuggable*) to denote if an application is a debug build. Figure 3 shows a part of an AndroidManifest.xml file with debugging enabled. This allows developers to perform tasks like collecting logs, etc.

It is a common practice to use self-signed or the same SSL certificate(s) on test servers during the debug phase and valid SSL certificates during deployment. To use these certificates developers have to either implement SSL pinning or completely disable SSL verification. Disabling verification is equivalent to accepting all SSL certificates.

The downside with adding code that disables SSL verification is that developers have to remove all traces of this code after debugging is complete. It is not uncommon to forget or leave active debug code in the deployed application. CWE (Common Weakness Enumeration) defines it as Leftover Debug Code [11] - *"The application can be deployed with active debugging code that can create unintended entry points"*. If there are any remnant instances of such debug code then the application would be vulnerable to MITM attacks.

Our solution to this problem is to link the debug flag in the application manifest with temporary removal of SSL verification. If the application is in *debug* build, then Android should accept any SSL certificate when creating an SSL connection. With our solution, the use of self-signed or untrusted CA certificates during debugging can be handled by Android itself. Thus, if the manifest has the debug flag set to *true*, Android can disable SSL verification, i.e. accept all SSL certificates. Once the debugging of the application is completed, the debug flag can be removed from the application manifest and Android can enforce SSL verification checks on all connections as it would normally do. Note, after debugging the developer should do one final test of the application in the release mode with the SSL certificates that are going to be used on the production servers to ensure correct functioning of the application.

In our solution, the developer does not have the need to actively add the debug flag to the application manifest. Android has implemented support for a true debug build starting with SDK Tools revision 8 (December 2010) [6]. This means the build tools add the attribute `android:debuggable` automatically to the application manifest. The SDK assumes all incremental builds to be debug builds, so it inserts `android:debuggable="true"` to the manifest. When exporting a signed release build, it removes the attribute. However, if the developer manually adds the debug flag to the manifest but forgets to remove it later from the production build, it is not automatically removed. We believe there is an easy solution to this i.e. enforcing a check at build time or at the application market level to see whether the debug flag is enabled and notifying respective developers to resubmit a corrected production build.

Thus, if SSL verification is linked with the debug flag, then developers will not have to worry about adding custom SSL verification code to use a particular certificate on test servers. This will alleviate the burden of keeping track of and removing all instances of extra SSL verification code from the application that they would have otherwise added.

### B. Allowing SSL Pinning as an Application Manifest Policy

SSL Pinning is a process of associating a particular certificate with a host or server. It allows developers to restrict the number of default root CA certificates as well as allows them to trust self-signed, invalid or untrusted CA certificates.

Android comes with a set of trusted root CA certificates that it uses to establish the chain of trust. Some versions of Android trust more than a hundred root CAs, for e.g., Android 4.2 trusts 140 root CA certificates. The developer might not want to trust all of them for reasons such as compromise of root CA credentials [8], [12]. In such cases, the developer can pin only a few root CAs she trusts. Developers also need to

```xml
<uses-SSLPinning useDefaultTrustStore="false" >
 <!--Self-signed Server Certificate-->
 <Cert type="self" algo="SHA-1">
    B8:01:...
 </Cert>
 <!--Trusted Issuer CA Certificate-->
 <Cert type="ca" algo="SHA-1">
    93:E6:...
 </Cert>
</uses-SSLPinning>
```

Listing 1. Implementing SSL Pinning as a policy in the *AndroidManifest.xml* file.

implement SSL pinning when they use self-signed or untrusted CA certificates.

In both cases, the developer would have to create a `KeyStore` containing those trusted certificates (self-signed or root CAs) and write custom SSL verification code to use this `KeyStore`, instead of the default used by Android. Another way of using self-signed or untrusted certificates in Android is to completely disable SSL verification i.e. *accept all certificates* which leaves the application vulnerable to MITM attacks.

Our solution to this problem is to enable the developer to pin SSL certificate(s) within the application manifest. This would eliminate the need for the developer to add custom SSL verification code or to disable SSL verification altogether. Listing 1 shows the *AndroidManifest.xml* for an application with our hypothetical SSL Pinning enabled in the manifest. The developer would add the trusted certificates as policies in the manifest. Note, a similar manifest policy for SSL pinning was also concurrently proposed by Fahl et al. [17] (cf. Section VI). However, they allow developers to only pin individual certificates to the manifest. Thus, if the developer wants to trust a particular root CA, she would have to implicitly trust all the intermediate CAs comprising the chain-of-trust. With our solution, we provide the `useDefaultTrustStore` option, which allows the developer to specify if she wants to trust only the pinned certificate or use it along with the default root CA certificate store from Android. If set to *true*, the application would treat both the pinned certificate and the default root CA certificate store of Android as trusted, otherwise it would trust only the pinned certificate.

Enabling SSL Pinning in the manifest can also tackle likely problems introduced by the use of DNS based load balancing techniques like GeoDNS [19]. GeoDNS routes the application's request to the closest server with respect to geo-location. In the event that certificates signed by untrusted CAs are used on some of the servers, the developer can pin these certificates in the manifest. As there is no definitive way of knowing beforehand which server the request will be redirected to, our solution allows the developer and service provider the flexibility of using any certificate on any server. This would prevent the need for disabling SSL verification, thus protecting the application from MITM vulnerabilities.

## IV. EVALUATION

We aim to show that developers need not add custom source code in the application for performing verification checks when using common SSL features. In order to do that we analyze the usage of SSL in the application along with the SSL certificate(s) used on their production servers.

### A. Experiment Methodology

**Determining the necessity for custom SSL verification:** The study by Fahl et al. [16] demonstrates the misuse of SSL in Android applications. However, their results do not distinguish between the vulnerabilities introduced by developer code and the those introduced because of the use of advertisement and analytics libraries (henceforth referred to as *ad* libraries). We wanted to focus on vulnerabilities caused due to custom verification code added by the developer. We believe that the SSL code generated due of the inclusion of ad libraries is orthogonal to the functionality of the application and does not affect it. Furthermore, if the vulnerable code is present in ad libraries, it can be easily addressed by having the library provider distribute an update. We classified applications that use SSL into three categories, namely:

- *src_only* - Applications that use SSL only in non-advertisement source i.e. no SSL code introduced because of ad libraries.
- *ads_only* - Applications which have SSL only because of inclusion of ad libraries.
- *src_and_ads* - Applications which have some SSL usage from non-advertisement code and some from ad libraries.

To classify applications in to the above mentioned categories, we generated a list of the most commonly used ad libraries. We obtained the top 100 libraries noted by Grace et al. [22] in their study of mobile advertisement library risks. To that list, we added the libraries obtained from AppBrain.com, a Website which shows the statistics for library usage in Android applications. While this list may not be exhaustive, from the statistics on AppBrain.com, the list of libraries that we obtained was present in 78.11% [5] of applications available in the Google Play Store, as of March 2013.

We examined applications that belonged to the *src_only* and *src_and_ads* categories. We disassembled each application using baksmali and used *grep* to search for keywords associated with SSL, enumerated in Table I, in the disassembled code and stored it along with the file path where SSL was used. We used a Python script to classify the usage of SSL into advertisement and non-advertisement code. A file containing the list of ad libraries was given to the Python script as input. Android application source, including ad libraries, follow a particular directory structure. We used this directory structure to determine if the SSL usage occurred in an ad library or in non-advertisement code. The directory path of the library contains the name of the library which we compared with the list of known ads and analytics libraries. The Python script classified an application as *src_only* if it used SSL only in non-advertisement code and as *src_and_ads* if it

TABLE I
SSL-RELATED KEYWORDS SEARCHED BY *grep*

```
SSLSocketFactory
TrustManager
HostnameVerifier
HttpsUrlConnection
SSLContext
SSLSocket
SSL
https://
```

used SSL in non-advertisement code as well as ad libraries. The applications which used SSL only in ad libraries were classified as *ads_only*.

The list of ads and analytics libraries that we used for classification is not exhaustive. Hence, it is possible that SSL code occurring from an ad library not in our list might be classified as non-advertisement SSL code (false negatives). Such applications would be classified as either *src_only* or *src_and_ads*. However, we analyzed applications belonging to both classes. In the applications where the SSL code was originating from a library, we examined whether the library was a developer tools library supplementing functionality or an ad library. This reduced the number of false negatives. We also analyzed the SSL code to determine whether it was used for creating custom SSL connections or not.

**Verifying SSL Certificate:**

We downloaded and verified the SSL certificates used on the production servers of the application to determine whether developer needed to add custom SSL verification code. We used a Python script to download the SSL certificate on the production server . To determine whether the SSL connection could be made using built-in APIs, we used *openssl* to verify each certificate using the default root CA store of Android. This enabled us to identify which applications needed custom SSL verification code for using self-signed or untrusted CA certificates and which did not. We used the root CA certificates from Android OS ver. 2.3.3, 4.0, 4.1, 4.2 because together they account for 94.3% of devices being used (as of May 2013 [3]).

We started with the analysis of open-source applications because Official Google Play Store applications are closed-source and only the binary (*.apk* file) is available.

### B. Analysis of Open-source Applications

We obtained open-source applications from F-droid, an Android Free and Open Source Software repository, that provides multiple versions of each app and its source. We downloaded a snapshot of all applications in the repository in January 2013 which consisted of 240 applications. We used *grep* to search each of these applications for terms associated with SSL (cf. Table I) and classified these applications into those using SSL and those that do not. Out of the 240 applications, we identified 26 that use SSL. None of the applications obtained from F-Droid repository contained ad libraries because of the repository policy. We manually analyzed the source code of these applications.

Out of the 26 applications, 10 contained insecure custom SSL verification code. Out of these 10, 6 applications used some form of a custom `TrustManager` that accepted all SSL certificates without verification. 3 applications used some form of a custom `HostnameVerifier` that allowed all certificates without performing hostname verification. One application used both techniques to accept all certificates and not verify server identity. Only 2 out of the 10 applications containing insecure custom SSL code, used self-signed or untrusted CA certificates on production servers. It is likely that this insecure code was added to enable the use of such certificates. The remaining applications were using valid SSL certificates and did not have the need for custom SSL verification code. Besides analyzing the code we also looked through the comments added by the developer. In the comments of one application, that used both the *Trust-All* and the *Allow-All* strategies (cf. Section IV-C), the developer mentioned that she had to use them because one of the third-party servers in her application used a self-signed certificate. However, the use of insecure SSL verification code for one service provider left all other service providers vulnerable to a MITM attack. To tackle such scenarios we suggest that the developer pin only (trusted) self-signed certificates and use them along with the default Android trusted CA certificate store, as opposed to completely disabling SSL verification.

The analysis of open-source applications was on too small a set of applications (26 out of 240 i.e. 10.83% applications used SSL) to generalize the behavior of all applications using SSL. The next logical step is to perform a large scale analysis of applications using SSL and see if they follow similar patterns.

### C. Patterns for insecure SSL verification

In some open-source applications, patterns pertaining to the custom SSL verification code looked remarkably similar to the patterns we found on StackOverflow.com entries. Rest of the applications used minor variations of these patterns. These code patterns, if present in the application, leave it vulnerable to MITM attacks. Following are the code patterns for insecure SSL verification code we found:

**No Certificate Validation a.k.a Trusting All Certificates:** The `TrustManager` interface in Android (*javax.net.ssl*) can be defined to implement custom certificate verification or extended validation features like SSL Pinning. It can be initialized to trust either Android's default root CA certificates or the CA certificates that the developer wants or both. Listing 2 shows the code idiom of the sample trust manager which trusts all certificates. The `checkServerTrusted()` function has to be overridden to define custom certificate verification. If it is defined with an empty code block then the trust manager does not perform any verification, accepts all SSL certificates and proceeds with the connection even if the certificate is invalid or expired. This trust manager is commonly referred to as *Trust-All Trust Manager*.

**No Hostname Verification a.k.a Allowing All Hostnames:** The `SSLSocketFactory` class (*javax.net.ssl*) uses a host-

```
1  TrustManager tm = new X509TrustManager(){
2
3   public void checkServerTrusted(X509Certificate[] c, String
      at) throws CertificateException
4    {} // <- Empty Code Block = No verification
5  };
```

Listing 2. A Trust Manager that trusts all certificates.

```
1  SSLSocketFactory s = new SSLSocketFactory(keyStore);
2  s.setHostnameVerifier(SSLSocketFactory.
    ALLOW_ALL_HOSTNAME_VERIFIER);
```

Listing 3. Using built-in Allow All Hostname Verifier

```
1  class IgnoreError implements WebViewClient{
2
3   public void onReceivedSslError(WebView w, SslErrorHandler
      h, SslError e) {
4     h.proceed(); //<- Proceed even if error occurs
5  }}
```

Listing 5. IgnoreWebViewClient which ignores SslError

name verifier for verifying the identity of the server. The default hostname verifier used by the built-in APIs performs all the necessary checks required. The Android SDK also provides an *Allow All Hostname Verifier*. However, it mentions in the documentation that this hostname verifier [2] neither performs hostname verification nor throws an exception if the hostname is invalid. It essentially turns hostname verification off. Listing 3 shows the code idiom for using a built-in *Allow All Hostname Verifier*.

The Android SDK also allows developers to define custom hostname verification in the `verify()` function of the `HostnameVerifier` interface. This function verifies the server hostname and throws an exception if it does not match with the certificate. However, if `verify()` is defined with an empty code block or if it always returns *true*, then the hostname verifier neither verifies the identity of the server nor throws an exception. Listing 4 shows the code idiom for such a hostname verifier. This hostname verifier is also referred as *Allow All Hostname Verifier*.

**Ignoring SSL Error:** A `WebView` is a `View` used in Android applications to allow browsing of Web pages. Developers can create custom SSL connections when browsing secure pages in the WebView. It uses the `WebViewClient` interface to perform handle SSL errors. The developer can override the `onReceivedSslError()` function in the `WebViewClient` to appropriately notify the user about the SSL error. However, this function can be defined such that the error handler ignores the SSL error and proceeds with the connection. Listing 5 shows the code idiom of such a `WebViewClient`.

### D. Analysis of Applications from Google Play Store

Our analysis of open-source Android applications depicted the presence of insecure custom SSL verification code even

```
1  class AllowAll implements javax.net.ssl.HostnameVerifier{
2   AllowAll() {
3      this.<init>();
4   }
5   public boolean verify(String r1, SSLSession r2) {
6      return true; //<- Always true = Allow All certificates
7  }}
```

Listing 4. A Hostname Verifier that allows all certificates

though a majority of the applications were using valid SSL certificates. We obtained 13,000 applications from the Official Google Play Store. We took a snapshot of the 500 most popular free applications from each of the 26 categories in the Google Play Store in March 2013. Since the source of these applications is not readily available, we disassembled the binary *apk* using baksmali. On the disassembled code, we used *grep* to search for keywords associated with SSL and identified 4,985 applications (38.34%) that used SSL.

*1) Manual Analysis of 200 applications:* We wanted to see if closed-source applications also contain the same code patterns for vulnerable SSL code. Hence, we randomly selected 200 applications for manual analysis. Another reason for manual analysis was to identify code patterns and use these patterns to categorize the remaining applications. We decompiled these applications using *Dare* [9]. *Dare* is a tool that retargets Android applications to *.class* files and then decompiles them into *.java* files using *Soot*. We used a Python script on the decompiled code to classify the applications based on where SSL was used. We looked at applications belonging to *src_only* and *src_and_ads* categories.

137 out of those 200 applications used SSL in non-advertisement code. 96 applications belonged to the *src_only* and the remaining 41 belonged to the *src_and_ads*. 84 out of 137 applications contained insecure SSL code which introduced MITM vulnerabilities. 58 applications used a `TrustManager` that did not perform any server certificate verification. 13 applications used a `HostnameVerifier` that did not perform hostname verification. 13 applications did neither server nor hostname verification.

For each of these applications, we also downloaded and verified the SSL certificates used on the production servers of the application using *openssl* and the default root CA store of Android. Surprisingly, only 5 out of the 84 applications using SSL, used self-signed certificates. The other 79 applications used valid SSL certificates and did not need custom SSL verification code. It is highly probable that the developer added custom SSL code to disable verification checks during the debugging phase to accommodate untrusted CA or self-signed certificates on development servers, but forgot to remove all instances of this code in production code. This confirmed our belief that closed-source applications also contain additional code for SSL verification that resembles the code patterns described in Section IV-C.

*2) Analysis of SSL verification code:* To determine the pervasiveness of the presence of code patterns in closed-source applications, we analyzed the disassembled code of the remaining applications that SSL. We classified these applications

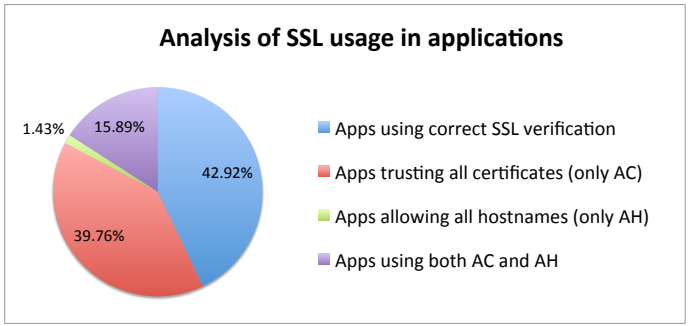

Fig. 4. Results for the analysis of SSL verification code with respect to the technique *Trust Any Certificate (AC)* or *Allow Any Hostname (AH)* used to bypass SSL verification.

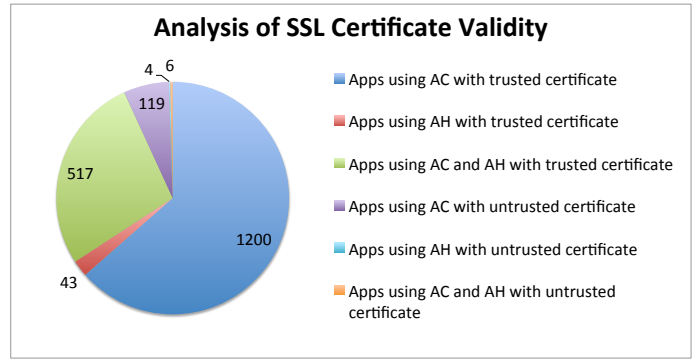

Fig. 5. Results for the analysis of SSL certificates used on the production servers of the application with respect to the technique used to bypass SSL verification i.e. *Trust Any Certificate (AC)* or *Allow Any Hostname (AH)*

based on where SSL was being used i.e. advertisement and non-advertisement code into the three categories mentioned above. 1,883 applications belonged to *src_only* while 1,282 applications belonged to *src_and_ads*, a total of 3,165 out of 4,785 applications.

We obtained a list of known classes which implement insecure custom SSL verification code patterns from Fahl et al. [16]. To this list, we added the classes that we identified as vulnerable during manual analysis of 200 applications and the analysis of open-source applications. We used this list to initially prune down the number of applications which have known classes containing insecure SSL verification code. For the remaining applications, we used *Androguard* [4] to obtain the source code of the classes which used SSL i.e. `TrustManagers`, `HostnameVerifiers` and `SSLSocketFactorys`, and analyzed them for presence of custom code that bypassed SSL verification.

943 applications contained known classes that bypass SSL verification. On manual analysis of the remaining applications, we found 680 more applications that contained some variation of the known classes that bypasses verification. Thus, 1,623 applications had these implementations of classes that bypassed SSL verification. We additionally found 182 applications that ignored SSL verification checks but their class names were obfuscated. In all, 1,805 out of the 3,165 applications contain custom SSL code that bypasses server certificate and hostname verification.

To see how many of these 1,805 applications actually need to use custom SSL code, we examined their SSL certificates. Only 86 applications used self-signed certificates, 3 applications used certificates signed by a CA not trusted by Android and 35 applications used expired certificates, which totals to 124. The remaining 1,681 applications had genuine valid certificates. They could have created SSL connections using one of the built-in APIs.

*E. Results*

**Applications that use SSL:** From the analysis of the SSL use in 4,985 closed-source applications obtained from Google Play Store, we identified 3,302 applications (137 from manual analysis and 3,165 from automated analysis) using SSL in non-advertisement code. Out of those 3,302 applications, 1,889 (57.20%) contain additional SSL code that bypasses verification. Figure 4 shows the breakdown of applications based on the strategy they use to bypass SSL verification. 39.94% (1,319) of applications only use some variation of a `TrustManager` that accepts all certificates (*Any Certificate, AC*) without verifying them, whereas 1.43% (47) of applications only use some variation of a `HostnameVerifier` that accepts any hostname (*Any Hostname, AH*) on the certificate irrespective of the server identity. Besides these, 15.89% (523) of applications accepted all certificates and allowed all hostnames. We can see that very few developers implement only custom hostname verification that allows any hostname, whereas a majority of them implement custom certificate verification that accepts any certificates. This can be attributed to the fact that the developer does not need to write any hostname verification code if they do not verify the certificate itself since the server hostname is verified after the certificate is verified. The remaining 42.8% of applications use SSL correctly and hence do not need to use our solutions. However, using our solutions could help the developers with an easier way to implement SSL features such as certificate pinning.

**Applications that use self-signed or untrusted CA certificates:** Out of the 1,889 applications having custom SSL code that bypasses verification, only 129 (3.90% of applications using SSL) applications used self-signed certificates or untrusted CA certificates. Figure 5 shows the results of SSL certificate analysis. Only 4 applications using a self-signed certificate on a production server use an Accept All `HostnameVerifier`; 119 applications use an Any Certificate `TrustManager` and 6 applications use both. It is likely that the developers of these applications added custom SSL code that bypasses verification to work with the self-signed or untrusted CA certificates. However, if SSL pinning is enabled as a configuration policy in the application manifest, developers could mention the certificates they want to trust in the manifest itself without having to add custom verification code to the application, thereby preventing such MITM vulnerabilities.

**Applications that use valid SSL certificate but bypass SSL**

**verification:** While the developers of the 129 applications (using self-signed or untrusted certificates) need to write custom verification code in order to use those certificates, the unnecessary presence of custom code that bypasses verification has left the remaining 1,760 (53.30%) applications vulnerable to MITM attacks. Figure 5 shows the number of applications having SSL code that bypasses verification and use valid certificates on their production servers. 1200 applications use an Any Certificate `TrustManager`; 43 applications use an Accept All `HostnameVerifier` and 517 applications use both, despite all applications using valid SSL certificates on the production servers. Since applications do not need custom SSL verification code to use valid certificates, the presence of this vulnerable code can be attributed to the developer adding that code during debugging and not removing all instances of that code from the deployed application. If the debug flag in the manifest is linked with SSL verification, the developer would not have to worry about adding custom SSL code for using untrusted CA or self-signed certificates on development servers.

Linking the debug flag to SSL verification and enabling SSL pinning as a configuration policy in the application manifest would be beneficial to at least these 1,889 applications as they remove the need for the developer to write additional custom verification code to the application. Our configuration based solutions could at the least prevent MITM vulnerabilities, introduced due to custom SSL code that bypasses verification, in 1,889 applications i.e. 57.20% of the applications using SSL in non-advertisement code.

## V. DISCUSSION

While our solutions involve both linking debugging using SSL certificates to the debug flag in the application manifest and enabling SSL pinning as a policy in the manifest, there could be some confusion if the developer wants to use both the solutions, for example, use the same self-signed certificate on test servers and production servers. In this case the developer would only need to pin the certificate. Since pinning would make the system treat it as a trusted certificate, the same certificate could be used for debugging on the development servers. The developer could also use both techniques to the same effect.

In Section III we provide a taxonomy of the major causes why developers add custom SSL verification code to the application, based on the study of entries related to SSL in the on-line forum StackOverflow.com. While the causes we provided are the predominant ones, the taxonomy might not be exhaustive. We believe that remaining possible causes constitute a minor percent of applications. There could be reasons such as insufficient and inconsistent API documentation causing developers to copy-paste code snippets from on-line forums. It could also be that the complexity of the built-in APIs drives the novice developer to use ready-made solutions without considering their consequences. In order to mitigate the vulnerabilities occurring due to these reasons, further investigation in this matter requiring co-operation from the developers is necessary.

## VI. RELATED WORK

Fahl et al. [16] perform static analysis on Android applications to show the widespread misuse of SSL. Concurrent to our work, follow on work by Fahl et al. [17] suggests solutions to tackle the problems concerning SSL misuse in Android applications. Fahl et al. suggested making SSL verification mandatory for developers; providing two new classes that replace the default Android trust manager; and implementing SSL pinning in the application manifest. We believe that making SSL verification mandatory and adding new classes for verification are invasive solutions and make customizing SSL connections even more complex. Our contribution is that a small change in the application manifest is sufficient to address the problem of developers having the need to add custom SSL verification code.

Georgiev et al. [20] have demonstrated how the lack of uniformity across the industry regarding standards for creating APIs has lead to a plethora of certificate validation bugs in commercial softwares and SDKs. They analyze various libraries and merchant SDKs, used in non-browser softwares and how incorrect implementation of certificate validation is leading to vulnerabilities in major software components. Sounthiraraj et al. [29] use static and dynamic analysis to enable the automatic, large-scale identification of SSLvulnerabilities in Android applications. They identify custom validation procedures in an application and extract information from it, which is used along with user interface enumeration to drive dynamic analysis on emulators.

SSL has been used as the de facto standard for securing communications over the Web but the current state of SSL leaves much to be desired for. A lot of research has looked into the implementation of SSL protocol and the flaws therein. Moxie Marlinspike has demonstrated how the infrastructure of SSL can be defeated because of the lack of basic constraint checking [23] and incorrect parsing of the *null* characters in the "CommonName" field of the certificate [24]. Percoco et al. [25] showed how fake SSL certificates can be used in mobile devices due to buggy certificate validation code present in mobile Web browsers. By contrast, we investigate malpractices while using SSL in Android applications.

There have been numerous efforts to investigate security vulnerabilities in the Android ecosystem such as the work by Porter Felt et al. [18], Enck et al. [13], Grace et al. [21], [22], Zhou et al. [31], Davi et al. [10], Bugiel et al. [7], Zhou et al. [1]. These studies show how the permission system of Android can be abused and how it can be prevented, but do not include the study of SSL in Android and its abuse. Our work highlights the vulnerabilities based on the incorrect use of SSL in Android apps. Since the permissions used by these apps during connection establishment are legitimate, the countermeasures provided in these studies would not prevent or mitigate the threats present due to incorrect use of SSL.

Several studies provide a good overview of the Android Security Model and its threat model, such as Enck et al. [14], Vidas et al. [30], Shabatai et al. [28]. In a comprehensive study on Android application security, Enck et al. [14] mention finding applications containing socket factories with names such as *TrustAllSSLSocketFactory* and *AllTrustSSLSocketFactory*. They discussed the presence of these socket factories as bearing a potential for vulnerabilities but could not find malicious use for them (cf. [14], Finding 13). Our work demonstrates the presence of SSL related vulnerabilities in Android applications and proposes solutions for the same.

## VII. CONCLUSION

We investigated Man-in-the-Middle SSL vulnerabilities in Android applications introduced because developers needed to add custom SSL code that bypasses verification in the application. We identified insufficient Android API support for commonly used SSL features such as pinning, as the driving reason for developers needing to write additional verification code for customizing SSL connections. We proposed that customization of SSL verification should be stated in an application's package manifest and provided two suggestions: 1) Linking SSL verification with the *debug* flag; and 2) Enabling SSL pinning in the application manifest. We studied a total of 13,240 applications from the F-Droid repository and official Google Play Store and evaluated the usefulness of our suggestions on a corpus of 3,302 applications that use SSL in non-advertisement code. We demonstrated that a total of 1,889 (57.20%) of 3,302 applications using SSL in non-advertisement code would benefit from implementing our solutions. Based on our findings, we observe that using application package configuration policies is an effective approach to mitigating Android application SSL vulnerabilities.

## ACKNOWLEDGEMENTS

This work was supported by NSF grant CNS-1222680. Any opinions, findings, and conclusions or recommendations expressed in this material are those of the authors and do not necessarily reflect the views of the NSF. We would also like to thank Adwait Nadkarni, Al Gorski, Ashwin Shashidharan, and the anonymous reviewers for their valuable feedback during the writing of this paper.